\documentclass[preprint,pre]{revtex4-2}
\usepackage{graphicx}
\usepackage{lipsum}
\usepackage{float}
\usepackage{amsmath}
\usepackage{amssymb}
\usepackage{mathtools}

\begin{document}

\title{Structure and dynamics in the low-density phase of a two-dimensional  
cellular automaton model of traffic flow}

\author{Gilad Hertzberg Rabinovich, Ofer Biham and Eytan Katzav} 
\affiliation{Racah Institute of Physics, The Hebrew University, Jerusalem 9190401, Israel}

\begin{abstract}

We analyze the structure and dynamics in
the low-density phase
of the deterministic two-dimensional cellular automaton model of traffic flow
introduced in
[O. Biham, A.A. Middleton and D. Levine, Phys. Rev. A 46, R6124 (1992)].
The model consists of
horizontally-oriented (H)
cars that move to the right and vertically-oriented (V) cars that move downward,
on a square lattice of size $L$
with periodic boundary conditions. 
Starting from a random initial state of density $p$, 
which is equally divided between the H and V-cars, 
the model exhibits a phase transition at a critical density $p_c$.  
For $p < p_c$ it evolves toward a free-flowing periodic (FFP) state, while for
$p > p_c$ it evolves toward a fully-jammed state or to an intermediate state of congested traffic.
In the FFP states, the H and V-cars segregate into 
homogeneous diagonal bands, in which 
they move freely without obstruction.
To analyze the convergence toward the 
FFP states 
we introduce a 
configuration-space distance measure 
$D(t) = D_{\parallel}(t) + D_{\perp}(t)$
between the state of the system at time $t$
and the set of FFP states.
The $D_{\parallel}(t)$ term accounts for the interactions between homotypic pairs of H (or V) cars,
while $D_{\perp}(t)$ accounts for the interactions 
between heterotypic pairs of H and V-cars.
We show that in the FFP states $D(t)=0$, while in all the other states $D(t) > 0$.
As the system evolves toward the FFP states,
there is a separation of time scales, where $D_{\parallel}(t)$ decays
very fast while $D_{\perp}(t)$ decays much more slowly.
Moreover, the time dependence of $D_{\perp}(t)$ is well fitted by
an exponentially truncated power-law decay of the form
$D_{\perp}(t) \sim t^{-\gamma} \exp ( - t/\tau_{\perp} )$,
where $\tau_{\perp}$ depends on $L$ and $p$. 
The power-law decay suggests avalanche-like dynamics with no characteristic scale,
while the exponential cutoff is imposed by the finite lattice size.

\clearpage
\newpage

\end{abstract}

\maketitle

\section{Introduction}

The statistical physics of traffic flow applies concepts from thermodynamics 
and nonlinear dynamics to questions such as how the interactions between individual cars
and drivers give rise to collective traffic patterns
\cite{Chowdhury2000,Helbing2001,Schadschneider2002,Nagatani2002}. 
By treating road traffic as a many-body system, researchers analyze phase transitions 
between different dynamical phases of traffic flow,
much like the transitions between phases of matter
\cite{Kerner2004,Kerner2009,Kerner2017}. 

The flow of highway traffic has been studied extensively using the 
Nagel-Schreckenberg (NS) model
\cite{Nagel1992}.
This is a CA model which is defined on a one-dimensional grid of $L$ cells, 
with open or periodic boundary conditions.
The model features integer velocities and a set of dynamical rules for acceleration, slowing-down
and randomization.
In spite of its simplicity, the NS model captures key features of real traffic, 
including spontaneous formation of traffic jams and phase transitions 
between free-flowing and congested traffic
\cite{Schreckenberg1995,Nagel1996}.
The randomization step is crucial in order to capture realistic traffic behavior
and to account for the irregularity of human behavior.
The NS model established a minimal microscopic framework for jam formation 
in highway traffic. Kerner and co-workers later extended this approach by introducing 
speed synchronization as a key behavioral ingredient 
\cite{Kerner1996,Kerner2002}.
They showed that this extension yields a generalized phase structure comprising 
free flow, synchronized flow, and wide moving jams, in agreement with 
empirical results.

Taking the deterministic limit of the NS model and restricting 
the velocities to $0$ (stopped) and $1$ (moving)
yields the simplest 1D traffic model, which is analogous to
CA rule 184 
\cite{Krug1988,Sasvari1997,Belitsky2005,Fuks2023} 
in Wolfram's numbering scheme
\cite{Wolfram1983,Wolfram1984}. 
Starting from a random initial state in which each cell is occupied
by a car with probability $p$ and empty with probability $1-p$, every
time step each car moves one step to the right unless the cell
in front is blocked by another car.
An important property of this model is that the number of cars is conserved
\cite{Boccara1998,Boccara2002}.
CA model 184 exhibits a phase transition at $p=1/2$, separating between a free-flowing phase
for $0 < p < 1/2$ in which the velocity of all the cars is $v=1$, and a partially jammed phase
for $1/2 < p < 1$ in which the average velocity is $\langle v \rangle = (1-p)/p$.
The transient behavior of CA rule 184 
in the low-density phase,
as it evolves toward a free-flowing state 
was studied in Refs.
\cite{Fuks1997,Fuks1999}.
Starting from a random initial state 
in the low-density limit of $p \ll 1$,  
the number of blocked cars at time $t$ decays exponentially like $p^t$.
This implies that in the low-density limit the convergence to a
free-flowing state is extremely fast.
The dynamics of CA rule 184 on finite lattices was recently studied in Ref. 
\cite{Jha2025}.
The relaxation times of the traffic jams were calculated, revealing characteristic
features of continuous phase transitions.

The analysis of more complex traffic systems, such as urban traffic networks,
requires two-dimensional CA models.
In particular, the deterministic model of Ref. 
\cite{Biham1992}, 
often referred to as the 
Biham-Middleton-Levine (BML) traffic flow model
\cite{Ishibashi1994,Tadaki1994,Nagatani1999,Ding2011},
simulates traffic flow in a grid-like city with cars moving in two perpendicular directions. 
The model exhibits a dynamical phase transition between free-flowing periodic (FFP) states
in the low-density phase
and fully-jammed states in the high-density phase.
In addition to the free-flowing states at low densities and the fully-jammed states
at high densities, just above the transition there is a narrow range of densities,
in which there are intermediate partially-jammed states
\cite{Dsouza2005,Olmos2015}.
These intermediate states exhibit coexistence between domains of 
free-flowing traffic and domains of jammed traffic.

Only a few rigorous mathematical results are known for the BML model.
The high-density limit was studied in Ref. 
\cite{Angel2005}
using methods of percolation theory.
It was proved that for any lattice size $L$, 
when the density $p$ is sufficiently close to $1$,
the model converges to a fully jammed state after a finite number of time steps.
In the opposite limit of low densities, it was shown that if the number of cars
is smaller than $L/2$, the system will self-organize into a free-flowing state
in which all the cars will move with velocity $1$
\cite{Austin2006}.
It was also shown that the critical density at which the transition from the
free-flowing phase to the fully-jammed phase takes place satisfies
$p_c < 1/2$
\cite{Chau1998}.

In this paper we analyze the structure and dynamics of
the FFP states that emerge in
the low-density phase of the BML traffic-flow model.
To this end, we introduce a configuration-space distance measure 
$D(t) = D_{\parallel}(t) + D_{\perp}(t)$,
between the state of the system at time $t$
and the set of FFP states.
The $D_{\parallel}(t)$ term accounts for the interactions between cars
of the same type, while $D_{\perp}(t)$ accounts for the interactions 
between cars of different types.
We show that in the FFP states $D(t)=0$, while in all other states $D(t) > 0$.
We analyze the time dependence of $D_{\parallel}(t)$ and $D_{\perp}(t)$ and 
find that there is a separation of time scales, where $D_{\parallel}(t)$ decays
very fast while $D_{\perp}(t)$ decays much more slowly.
It is also shown that the time dependence of $D_{\perp}(t)$ is well fitted by
an exponentially-truncated power-law decay
of the form
$D_{\perp}(t) \sim t^{-\gamma} \exp ( - t/\tau_{\perp} )$,
where the power-law exponent is $\gamma \simeq 1.1$ and the relaxation time
$\tau_{\perp}$ depends on the lattice size $L$ and on the density $p$.
 
The paper is organized as follows.
In Sec. II we review the BML model.
In Sec. III we present a matrix formulation of the dynamical rules.
In Sec. IV we characterize the FFP states that emerge
in the low density phase.
In Sec. V we introduce the distance measure $D(t)$.
In Sec. VI we use the distance measure $D(t)$ to analyze the convergence
of the system toward the FFP states.
The results are discussed in Sec. VII and summarized in Sec. VIII.
In Appendix A we show that matrix formulation satisfies the
closure condition over the field of binary numbers.
In Appendix B we show that once the distance satisfies $D(t)=0$
at time $t$, it will remain zero at any later time.

\section{The Model}

The BML traffic flow model 
\cite{Biham1992}
is defined on a square lattice consisting 
of $L \times L$ cells,
such that each cell may contain at most one car at any given time.
The lattice exhibits periodic boundary conditions in both directions,
such that the square lattice is embedded in a torus. 
The model includes two types of cars: horizontally-oriented  (H) cars that move to the
right along the rows of the lattice and vertically-oriented (V) cars that move 
downward along the columns of the lattice.
It is convenient to identify the cells of the lattice with the matrix 
elements $(i,j)$ of an $L \times L$ matrix,
where $i,j=0,1,\dots,L-1$.
In this notation, an H-car at cell $(i,j)$ moves into cell
$(i, j+1 \ {\rm mod} \ L)$, while a V-car at cell $(i,j)$
moves into $(i+1 \ {\rm mod} \ L ,j)$.
The modular notation accounts for the periodic boundary conditions.
This means that an H-car located in the rightmost column moves from 
$(i,L-1)$ to $(i,0)$, while a V-car located in the bottom row moves from 
$(L-1,j)$ to $(0,j)$.
In the rest of the paper, for simplicity, we drop the ${\rm mod} \ L$ notation from the indices $(i,j)$,
assuming that in all cases $i,j = 0,1,2,\dots, L-1$ and that the modulo operation is taken
whenever needed to bring the indices $i$ and $j$ into this range.

In the BML model, the dynamics is controlled by a global traffic light, 
such that at each time step cars of one
type are allowed to move while cars of the other type remain in place.
This perfect synchronization of the signal timing is a simplifying assumption of the model.
Without loss of generality, we make the choice that starting from the random initial
state at time $t=0$,
at odd time steps each H 
car moves one step to the right, unless the cell on the right is occupied.
Similarly, at even time steps each V-car moves one step downward,
unless the cell below it is occupied.
In some cases it will be convenient to denote 
the odd time steps by $t = 2 s + 1$ 
and the even time steps by $t = 2 s$.
In this notation, the variable $ s = \lfloor t / 2 \rfloor $,
where $\lfloor x \rfloor$ is the integer part of $x$,
counts the number of double-steps,
where each double-step includes a move of the H-cars followed by a 
move of the V-cars.
Another time scale that we use below is the cycle, which consists of $2L$ time 
steps, or $L$ double-steps.
The cycle is the time it takes for
an H-car or a V-car that moves without obstruction to return to its initial cell.
The advantage of the cycle is that it provides a universal time scale that is
invariant to the system size.
In particular, the period of the FFP states is one cycle.
When we refer to the configuration at time $t$, it means the configuration obtained 
after the step of time $t$ has been performed.

In the initial state at time $t=0$
the cars are distributed uniformly at random with density $p$.
Here we focus on the case in which the densities of the H and
the V-cars are equal.
In this case, the
initial state is prepared by inserting in each cell
an H-car with probability $p/2$ or a V-car with probability $p/2$,
otherwise leaving it empty with probability $1-p$.
The states of different cells at time $t=0$ are thus uncorrelated.

The dynamics of the BML model satisfies a large number of conservation laws.
For example, the numbers $N_{\rm H}$ and $N_{\rm V}$ of H-cars and V-cars
are conserved in each random instance of the model.
Moreover, the number of H-cars in each row remains fixed 
and the number of V-cars in each column remains fixed.
Since the system is discrete and deterministic,
the initial state at time $t=0$
fully determines the outcome at any later time.
Since the lattice size is finite, 
the dynamics must eventually converge either to 
a fixed point (in which all the cars get stuck) 
or to a periodic cycle in which the system returns to the same
state every $T$ time steps.
It is important to note that we consider the cars as distinguishable
particles, so the period $T$ of the system is not merely the time after which the
configuration returns to its earlier form, but the time at which 
every single car returns to its earlier position.
Under this restriction, the shortest possible period of the system is of
one cycle or $T=2 L$ time steps.

\section{Matrix formulation of the dynamical rules}

The state of the system at time $t$ can be described in 
terms of two time-dependent $L \times L$
binary matrices, $H_t(i,j)$ and $V_t(i,j)$, 
$i,j = 0,1,2,\dots,L-1$
\cite{Bapat2014},
which account for 
the locations of the H-cars and the V-cars, respectively. 
More specifically, $H_t(i,j)=1$ in case that the $(i,j)$ cell
is occupied by an H-car at time $t$,
otherwise $H_t(i,j)=0$.
Similarly $V_t(i,j)=1$ in case that the $(i,j)$ cell is 
occupied by a V-car at time $t$,
otherwise $V_t(i,j)=0$.
Clearly, a cell cannot be occupied simultaneously 
by an H-car and by a V-car.
This constraint is expressed by the condition

\begin{equation}
H_t(i,j) V_t(i,j) = 0 
\label{eq:no_overlap}
\end{equation}

\noindent
for all pairs of 
$i,j = 0,1,2,\dots,L-1$.

The random initial state, at time $t=0$, is constructed as follows: 
for each matrix element $(i,j)$,
with probability $p/2$ we set $H_0(i,j) = 1$ and $V_0(i,j)=0$,
with probability $p/2$ we set $H_0(i,j) = 0$ and $V_0(i,j)=1$,
and with probability $1-p$ we set $H_0(i,j)=V_0(i,j)=0$.
Starting from such random initial state 
at time $t=0$,
the dynamics of the system at odd time steps, 
in which the H-cars move,
can be described by the evolution equations 

\begin{eqnarray}
H_{2 s +1}(i,j) &=& H_{2s}(i,j-1) \left[ 1 - H_{2s}(i,j) - V_{2s}(i,j) \right]
\nonumber \\
&+& H_{2s}(i,j) \left[ H_{2s}(i,j+1) + V_{2s}(i,j+1) \right]
\nonumber \\
V_{2s+1}(i,j) &=& V_{2s}(i,j).
\label{eq:Hupdate}
\end{eqnarray}

\noindent
Similarly, the dynamics at even time steps, in which the V-cars move,
can be described by the equations

\begin{eqnarray}
H_{2s+2}(i,j) &=& H_{2s+1}(i,j)
\nonumber \\
V_{2s+2}(i,j) &=& V_{2s+1}(i-1,j) \left[ 1 - H_{2s+1}(i,j) - V_{2s+1}(i,j) \right]
\nonumber \\
&+& V_{2s+1}(i,j) \left[ H_{2s+1}(i+1,j) + V_{2s+1}(i+1,j) \right].
\label{eq:Vupdate}
\end{eqnarray}

\noindent
The term in the first line on the right hand side of Eq. (\ref{eq:Hupdate}) accounts for the
motion of an H-car from cell $(i,j-1)$ at time $t=2s$ to cell $(i,j)$ at time $t=2s+1$.
Such step takes place in case that the $(i,j-1)$ cell is occupied and the $(i,j)$ cell is unoccupied
at time $t=2s$.
The term in the second line of Eq. (\ref{eq:Hupdate}) accounts for the case in which
the cell $(i,j)$ is occupied by an H-car at time $t=2s$ and this car is blocked 
because the cell $(i,j+1)$ is also occupied, by either an H-car or a V-car.
The third line in Eq. (\ref{eq:Hupdate}) accounts for the V-cars, which do not
move at odd time steps.
Eq. (\ref{eq:Vupdate}) is analogous to Eq. (\ref{eq:Hupdate}), replacing the roles of
the H-cars and the V-cars and the roles of the rows and the columns.

The evolution equations (\ref{eq:Hupdate}) and (\ref{eq:Vupdate}) may also be 
written in the form of Boolean equations, using the operators AND, OR and NOT
\cite{Givant2009}.
However, since we later use them within non-Boolean expressions, we found it
preferable to write these equations in an algebraic form.
As a result, we need to show that these equations 
are consistent, namely that they
satisfy the closure 
condition over the binary field and that the
condition of Eq. (\ref{eq:no_overlap}) is maintained.
These properties are shown in Appendix A.

For the analysis below it is useful to consider the 
diagonals of the $L \times L$ lattice.
More specifically, we focus on the counter-diagonals, which are parallel to 
the diagonal that runs from the lower-left
corner $(L-1,0)$ to the upper-right corner $(0,L-1)$. 
These diagonals can be labeled by $n=0,1,\dots,L-1$, where the diagonal $n$
consists of all the cells whose coordinates $(i,j)$ satisfy 
$ i+j = n \ ( {\rm mod} \ L)$.
The number of H-cars in the cells of diagonal $n$ at time $t$ is given by

\begin{equation}
h_t(n) = \sum_{i=0}^{L-1} H_t(i, n-i+L).
\label{eq:h_t_n}
\end{equation}

\noindent
Similarly, the number of V-cars in the cells of diagonal $n$ at time $t$ is

\begin{equation}
v_t(n) = \sum_{i=0}^{L-1} V_t(i, n-i+L).
\label{eq:v_t_n}
\end{equation}

\noindent
The time dependence of $h_t(n)$ and $v_t(n)$ can be obtained by applying
the evolution equations (\ref{eq:Hupdate}) and (\ref{eq:Vupdate}) on
the right hand sides of Eqs. (\ref{eq:h_t_n}) and (\ref{eq:v_t_n}).
To explore the evolution of these diagonals it is often convenient to use
a convective reference frame that goes along with the moving cars.
Focusing on the even time steps, in the convective reference frame
we express the contents of the diagonals in the form
$h_{2s}^{c}(n) = h_{2s}(n+s)$
and
$v_{2s}^{c}(n) = v_{2s}(n+s)$.
In the low density phase, as time evolves, some of the diagonals become
dominated by H-cars, other diagonals become dominated by V-cars 
and yet other diagonals become empty.

\section{The free-flowing periodic states}

Starting from a random initial state, in the low density phase, the system evolves 
toward an FFP state, in which the flow is free and uninterrupted.
%Below we refer to these states as free-flowing periodic (FFP) states.
The structures that emerge in the FFP states
exhibit segregated diagonals such that some diagonals
contain only H-cars, some contain only V-cars and in other diagonals all 
the cells are empty.
In these structures there is no interference between cars, such that all cars can move
forward at full speed.
Since the dynamical rules are deterministic, the FFP states are periodic
with period $T=2L$.
This implies that each car will return to its present position after $2L$ time steps,
or $H_{t+2L}(i,j) = H_{t}(i,j)$
and 
$V_{t+2L}(i,j) = V_{t}(i,j)$.
Once the system enters an FFP state, it remains permanently trapped in
this particular cycle.
The FFP states can thus be considered as absorbing states of the BML traffic model
\cite{Hinrichsen2000}.
The term FFP state refers to the entire cycle rather than to an instantaneous
configuration of the system.
In this sense, the FFP states in the low-density phase of the BML model 
are different from the absorbing states in other systems,
which are typically static states
\cite{Vespignani2000,DallAsta2008}.

The FFP states exhibit several features:

\begin{enumerate}

\item
There are no pairs of adjacent H-cars in the same row,
namely $H_t(i,j) H_t(i,j+1) = 0$ for all $i,j=0,1,\dots,L-1$.

\item
There are no pairs of adjacent V-cars in the same column,
namely $V_t(i,j) V_t(i+1,j) = 0$ for all $i,j=0,1,\dots,L-1$. 

\item
Each counter-diagonal $n=0,1,2,\dots,L-1$ 
may contain either H-cars or V-cars but not both,
namely if 
$h_t(n) \ge 1$
then
$v_t(n) = 0$
and vice versa.

\item
If at time $t=2s$ ($t=2s+1$) a counter-diagonal $n$ contains one or more H (V) cars
(which are about to move in the next time step),
then the next counter-diagonal $n+1$
cannot contain any V (H) cars,
because their existence would obstruct the cars behind them.
This implies that at even times 
$h_{2s}(n) v_{2s}(n+1) = 0$
for $n=0,1,\dots,L-1$.
Similarly, at odd times 
$v_{2s+1}(n) h_{2s+1}(n+1) = 0$
for $n=0,1,\dots,L-1$.

\end{enumerate}

To explore the range of densities in which FFP states exist, consider 
the case in which $N_{\rm H} = N_{\rm V}$.
Under this condition, FFP states exist in the range of
$0 < N_{\rm H}, N_{\rm V} \le L^2/3$
(assuming that $L$ is divisible by 3).
The densest possible configuration, at even times $t=2s$, 
consists of a periodic sequence of counter-diagonals,
in which the first diagonal is fully occupied by V-cars, 
the second diagonal is fully occupied by H-cars and the
third diagonal is empty.
The empty diagonal enables all the H-cars to move to the right and
make room for the V-cars to move downward in the next time step.

\begin{figure}
\includegraphics[width=12.0cm]{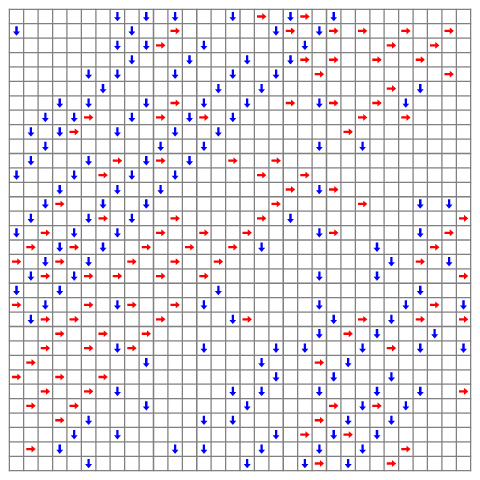} 
\caption{
(Color online)
Illustration of a free-flowing periodic state of the BML traffic flow model, 
obtained from a computer simulation of the model starting from a random initial state,
for a lattice of size $L=32$ and density $p=0.25$.
In this snapshot the time $t$ is even, such that in the next time step it will be
the turn of the H-cars to move. 
One can thus observe that the cells in front of all the H-cars are empty,
while the cells in front of some of the V-cars are occupied by H-cars.
In this configuration and in subsequent time steps,
all the cars move freely when their turn arrives,
with no obstructions.
}
\label{fig:1}
\end{figure}

In Fig. \ref{fig:1} we present an 
illustration of an FFP state of the BML traffic flow model, obtained from a simulation 
starting from a random initial state,  
for a lattice of size $L=32$ and density of $p=0.25$.
In this snapshot the time $t$ is even, such that in the next time step it will be
the turn of the H-cars to move. 
One can thus observe that the cells in front of all the H-cars are empty,
while the cells in front of some of the V-cars may be occupied by H-cars.
In this configuration all the cars will move freely without obstruction in the
next time step and in all future time steps.

It turns out that for small system sizes, in the low density phase, 
starting from a random initial state the model does
not always converge to an FFP state. In some cases it may converge to 
states we refer to as almost FFP states, in which at each time step there
are still a handful of blocked cars. These states, which persist for very long times,
appear to be periodic states with very long periods.
To test the convergence to FFP states, we carried out simulations of the model
with density $p=0.25$ and system sizes of $L = 32$, $64$, $128$, $256$, $512$ and $1,024$.
For each system size we simulated $1,000$ independent
instances with random initial states.
Each run was carried out for 100 cycles, which is much longer than the typical
time it takes to reach an FFP state.
For each lattice size we recorded the number of instances in which the model
did not converge to an FFP state during the simulation time.
It was found that for $L=32$ there were 190 instances that did not converge
to an FFP state, for $L=64$ there were 50 such instances, for $L=128$ there were
24 such instances and for $L=256$ there were 6 such instances.
In contrast, for $L=512$ and $1,024$ all the $1,000$ instances converged to 
FFP states within the simulation time.
This implies that for small systems up to the range of $L=256$, finite size
effects may hamper the convergence to FFP states.
For sizes in the range of $L=512$ and above, the model enters the large
system limit, in which instances that do not converge to an FFP state are
extremely unlikely.

Below we introduce a measure $D(t)$ of the distance between the configuration
of the system at time $t$ and the set of FFP states.
An important property of this distance measure is that $D(t)=0$ for any FFP state
and $D(t) > 0$ for any other state of the system.
Therefore, if $D(t)=0$ at a given time $t$ it implies that
the system has already converged to an FFP state
and that the traffic will continue to flow freely at any later time.
The distance measure $D(t)$ will enable us to analyze the convergence of the system toward
the FFP states.

\section{The distance measure}

Using these notations, we derive a formal condition 
that will enable us to check whether 
a given configuration, given by the matrices $H_t(i,j)$ and $V_t(i,j)$ is an FFP state.
To this end we introduce the distance measure

\begin{equation}
D(t) = D_{\parallel}(t) + D_{\perp}(t),
\end{equation}

\noindent
where $D_{\parallel}(t)$ 
accounts for the interactions between homotypic pairs of H (or V) cars,
while $D_{\perp}(t)$ accounts for the interactions between 
heterotypic pairs of H and V cars.

To properly calibrate these distance measures, we choose a normalization
condition under which for any system size $L$ and for any density $p$,
the expectation values of both measures for a random initial state at
time $t=0$ satisfy
$\langle D_{\parallel}(0) \rangle = \langle D_{\perp}(0) \rangle = 1$.
To this end, we express the measure $D_{\parallel}(t)$ in the form

\begin{equation}
D_{\parallel}(t) = \frac{ d_{\parallel}(t) }{ \langle d_{\parallel}(0) \rangle },
\end{equation}

%The distance measure $d_{\parallel}(t)$ is given by

\noindent
where

\begin{equation}
d_{\parallel}(t)   =    
\sum_{i,j=0}^{L-1}  H_t(i,j) H_t(i,j+1)
+  
\sum_{i,j=0}^{L-1}   V_t(i,j) V_t(i+1,j),
\label{eq:D_par}
\end{equation}

\noindent
and $\langle d_{\parallel}(0) \rangle$ is the expectation value of
$d_{\parallel}(t)$ at time $t=0$.
The first term
on the right hand side of Eq. (\ref{eq:D_par})
counts the number of pairs of H-cars that are adjacent to each other in the
horizontal direction, such that the car in front 
blocks the motion of the car in the back.
Similarly, the second term counts the number of 
pairs of V-cars that are adjacent to each
other in the vertical direction, such that the car in 
front blocks the motion of the car in the back.
In case that the car density is $p$, in the uniformly random initial state at $t=0$,
the expectation value of $d_{\parallel}(0)$ is given by
$\langle d_{\parallel}(0) \rangle = L^2 p^2/2$,
which implies that

\begin{equation}
D_{\parallel}(t) = \frac{2}{ (Lp)^2 } d_{\parallel}(t).
\label{eq:D_parallel_n}
\end{equation}

\noindent
We also express the measure $D_{\perp}(t)$ in the form

\begin{equation}
D_{\perp}(t) = \frac{ d_{\perp}(t) }{ \langle d_{\perp}(0) \rangle },
\end{equation}

\noindent
where

\begin{eqnarray}
d_{\perp}(t)  &=&  
\sum_{n=0}^{L-1} \min[h_t(n),v_t(n)]
\nonumber \\
&+& [ 1 - \pi(t) ] \sum_{n=0}^{L-1} \min[h_t(n),v_t(n+1)] 
\nonumber \\
&+& \pi(t) \sum_{n=0}^{L-1} \min[v_t(n),h_t(n+1)],
\label{eq:D_perp}
\end{eqnarray}

\noindent
and $\langle d_{\perp}(0) \rangle$ is the expectation value of
$d_{\perp}(t)$ at time $t=0$.
The function $\pi(t)$,
on the right hand side of Eq. (\ref{eq:D_perp}),
is the parity function,
which is given by

\begin{equation}
\pi(t) =  \left\{ 
\begin{array}{ll}
1 &   \ \ \     $t$ \ \ {\rm odd}
\\
0 &   \ \ \     $t$ \ \ {\rm even}.
\end{array}
\right.
\label{eq:parity}
\end{equation}

\noindent
The first term on the right hand side of Eq. (\ref{eq:D_perp}) is a
sum over all the counter-diagonals of the number of minority cars
in each counter-diagonal. In case that a counter-diagonal contains only H-cars
or only V-cars, its contribution to the first term is zero.
The second term,
which contributes only at even time steps,
is a sum of the minima between the number of H-cars $h_t(n)$ in counter-diagonal $n$ and the number 
of V-cars $v_t(n+1)$ in counter-diagonal $n+1$.
In case that either $h_t(n)$ is zero or $v_t(n+1)$ is zero, 
the contribution to the second term vanishes.
The third term,
which contributes only at odd time steps,
is a sum of the minima between the number of V-cars $v_t(n)$ in counter-diagonal $n$ and the number 
of H-cars $h_t(n+1)$ in counter-diagonal $n+1$.
In case that either $v_t(n)$ is zero or $h_t(n+1)$ is zero, 
the contribution to the third term vanishes.
The expectation value of $d_{\perp}(0)$ is given by
$\langle d_{\perp}(0) \rangle = L^2 p$,
which implies that

\begin{equation}
D_{\perp}(t) = \frac{1}{L^2 p} d_{\perp}(t).
\label{eq:D_perp_n}
\end{equation}

To gain insight into the geometric properties of $d_{\perp}(t)$,
given by Eq. (\ref{eq:D_perp}),
we consider a vector of length $2L$,
in which  the first $L$ components
are given by
$\min[h_t(n),v_t(n)]$, $n=0,1,\dots,L-1$.
The last $L$ components of this vector are given by
$\min[h_t(n),v_t(n+1)]$, $n=0,1,\dots,L-1$
(at even time steps)
or by
$\min[v_t(n),h_t(n+1)]$ 
(at odd time steps). 
The measure $d_{\perp}(t)$ 
is in fact the Hamming distance 
\cite{Hamming,Deza2016}
of this $2L$-dimensional vector from the origin.

The distance measure $d(t)$ quantifies the distance between the state of the system
at time $t$ and the set of FFP states.
When $d(t)$ is large the system is far away 
from the set of FFP states, while
as $d(t)$ decreases, the system gets 
closer to an FFP state.
In case that $d(t)=0$  
the system has reached an FFP state.
In Appendix B we show that if $d(t)=0$
then also $d(t+1)=0$, which implies that
once the system has reached an FFP state 
all the cars will continue to move freely with
no obstruction in all future time steps.

\section{Convergence to the free-flowing periodic states}

Below we analyze the self-organization processes in which a system that
starts from a random initial state converges toward an FFP state.
These processes take place at several levels: 

\begin{enumerate}

\item
Spreading of H (V) cars along the rows (columns): 
At the level of rows and columns, the H-cars in each row and
the V-cars in each column gradually spread and 
redistribute more evenly. 
The increased separation between the cars reduces the frequency 
of car blockages.
This is as if cars of the same type experience mutual repulsion, like charged particles,
causing them to redistribute more uniformly along a row or a column.

\item
Intra-diagonal segregation: 
The counter-diagonals gradually segregate such that each counter-diagonal 
is dominated by either H-cars or by V-cars.

\item
Inter-diagonal segregation and band formation:
The counter-diagonals dominated by H-cars and those dominated by V-cars segregate
into homogeneous diagonal bands that consist of a single type of cars.

\item
Pattern formation inside the counter-diagonal bands:
The bands form patterns in which the counter-diagonals at the front are
denser. The cars at the front of a counter-diagonal band tend to form a checker-board structure of 
occupied and empty cells, while the counter-diagonals at the back
are more disordered.

\end{enumerate}

In the case of interactions between cars of the same type, 
each car may interact directly only with two other cars, namely
the car in front of it and the car behind it.
The interaction is directional in the sense that each car may only affect the
motion of the car behind and may be affected by the car in front.
The H-cars in different rows and the V-cars in different columns
do not interact with each other directly. They interact only indirectly via the
cars of the other type. 

The interaction between cars of different types is more complicated.
It takes place along the ascending diagonals.
To illustrate this point, consider a configuration in which only the $n$th diagonal,
which consists of the cells $(i,n-i+L)$, $i=0,1,\dots,L-1$,
is occupied, and contains both H-cars and V-cars.
The clock is set to $t=0$ such that at the next time step it is the turn of the H-cars to move.
Under such conditions, each V car will later encounter the nearest 
H-car on its lower-left side.
Upon encountering each other the V-car will be blocked for one time step.
Eventually, all the V-cars will fall behind and will lag one step behind the
H-cars.

This example illustrates the segregation process, in which at the level of a single diagonal, after
one cycle of double-steps it becomes fully segregated. In this process, the cars that move
first in the cycle push all the cars of the other type to the adjacent diagonal in the back side.
In fact, the time it takes to complete the segregation process of a single diagonal may be 
shorter than a complete cycle. It is determined by the maximal distance between a 
V-car and the nearest H-car in the diagonal on its (lower) left hand side.
The segregation process makes the diagonals homogeneous, where some diagonals
contain only H-cars and other diagonals contain only V-cars.
In addition, there are empty diagonals which serve as buffers and enable the occupied
diagonals to move forward at their turn.

\begin{figure}
\includegraphics[width=7.5cm]{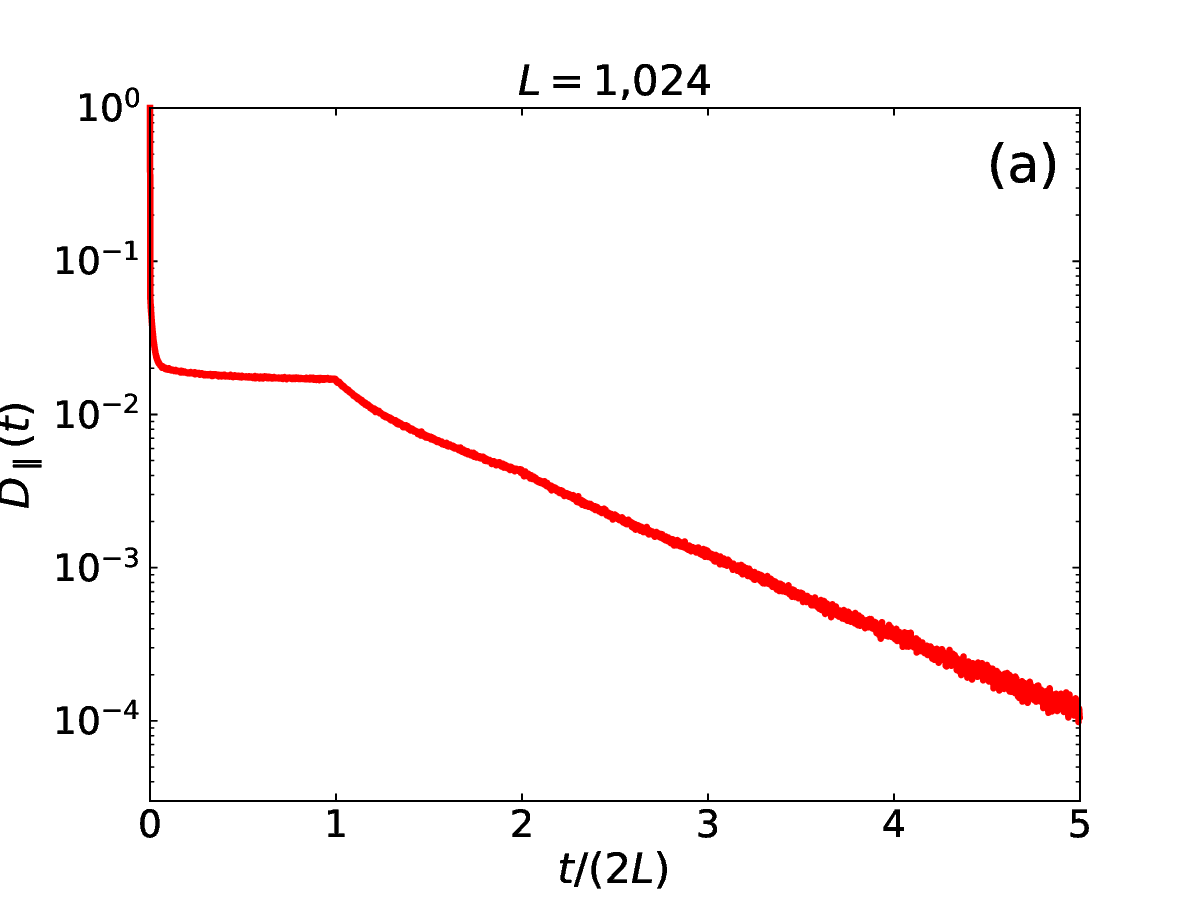} 
\includegraphics[width=7.5cm]{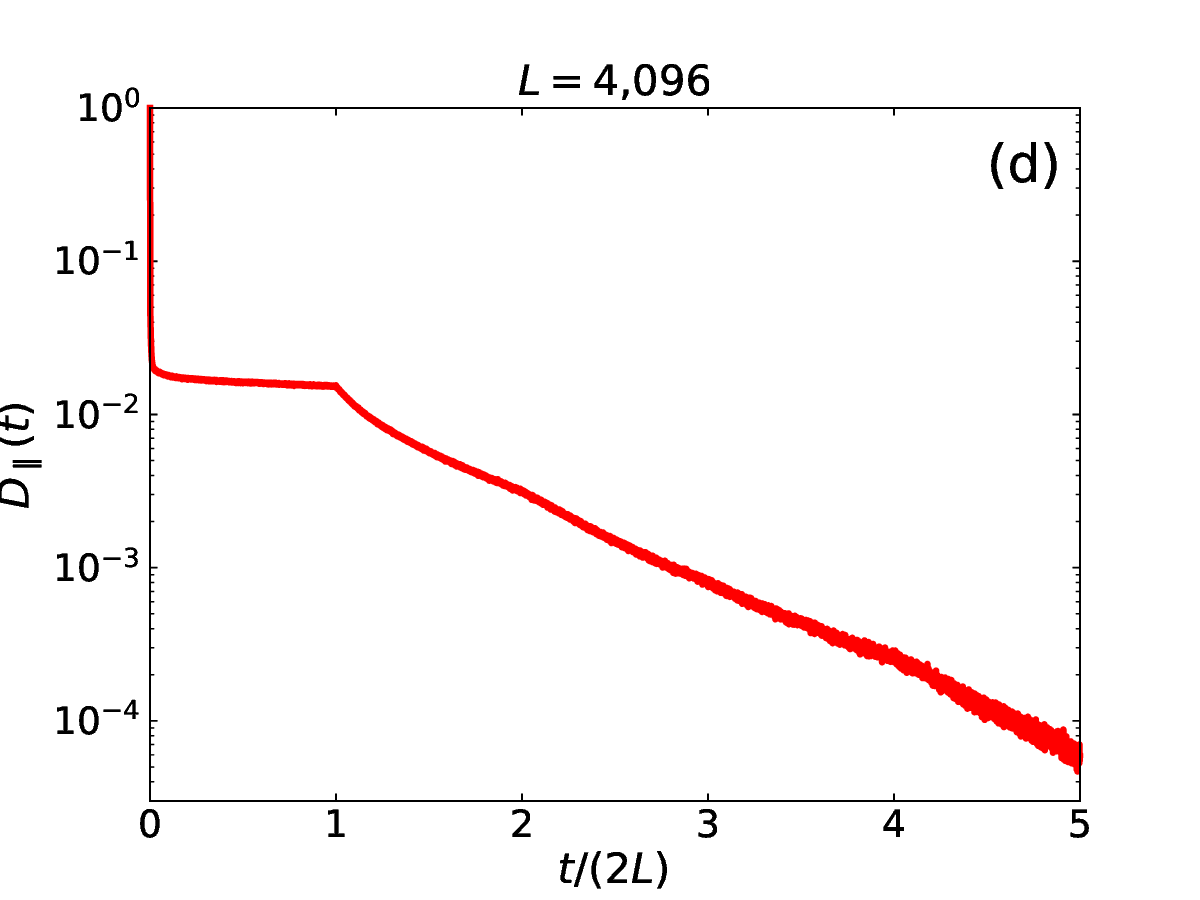} 
\\
\includegraphics[width=7.5cm]{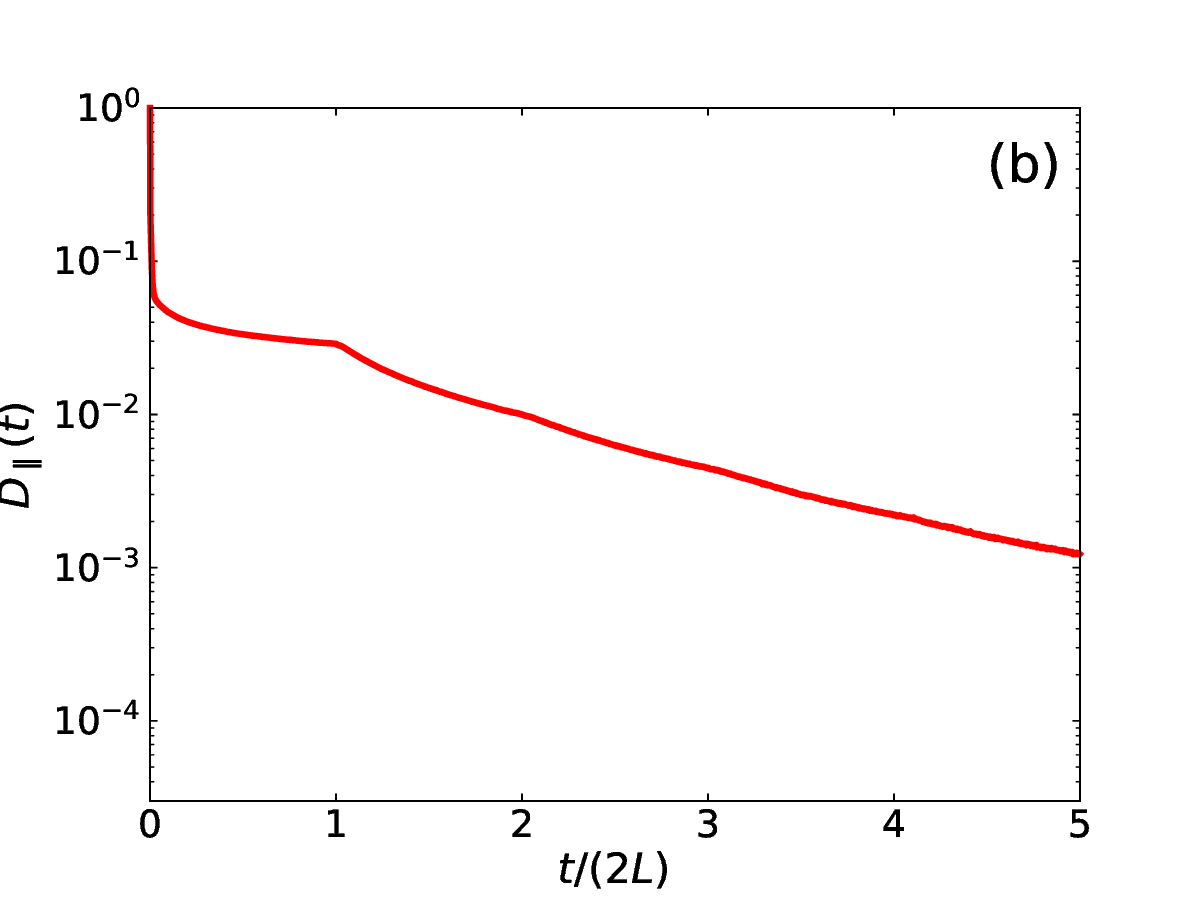} 
\includegraphics[width=7.5cm]{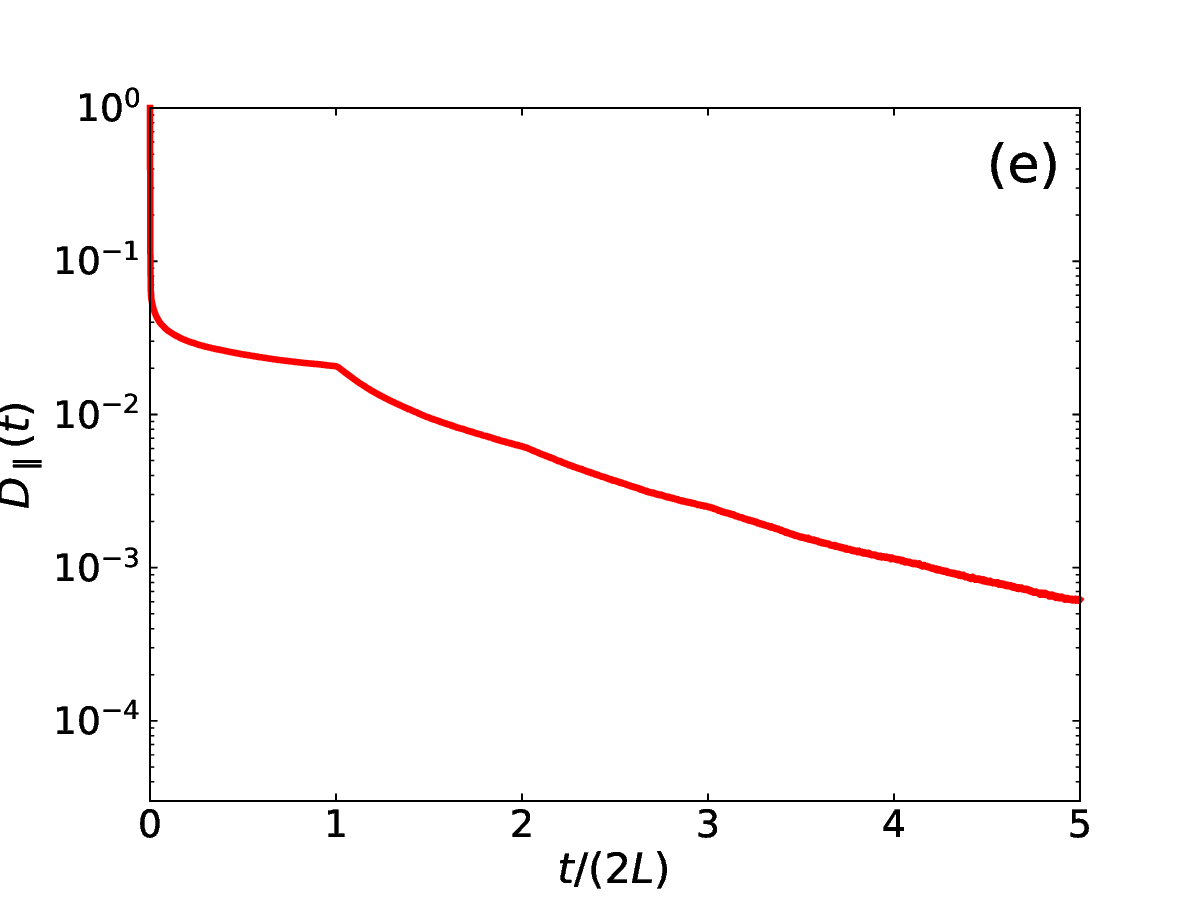} 
\\
\includegraphics[width=7.5cm]{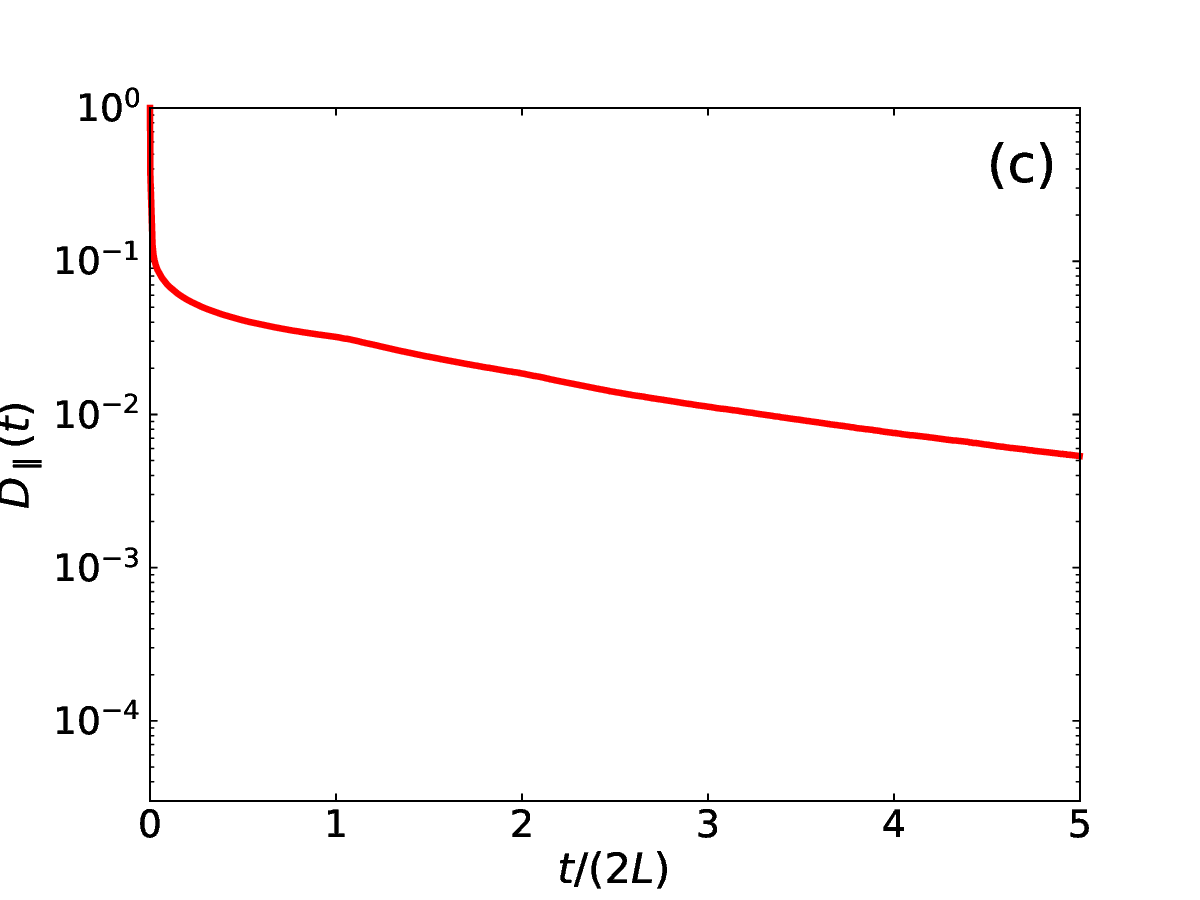} 
\includegraphics[width=7.5cm]{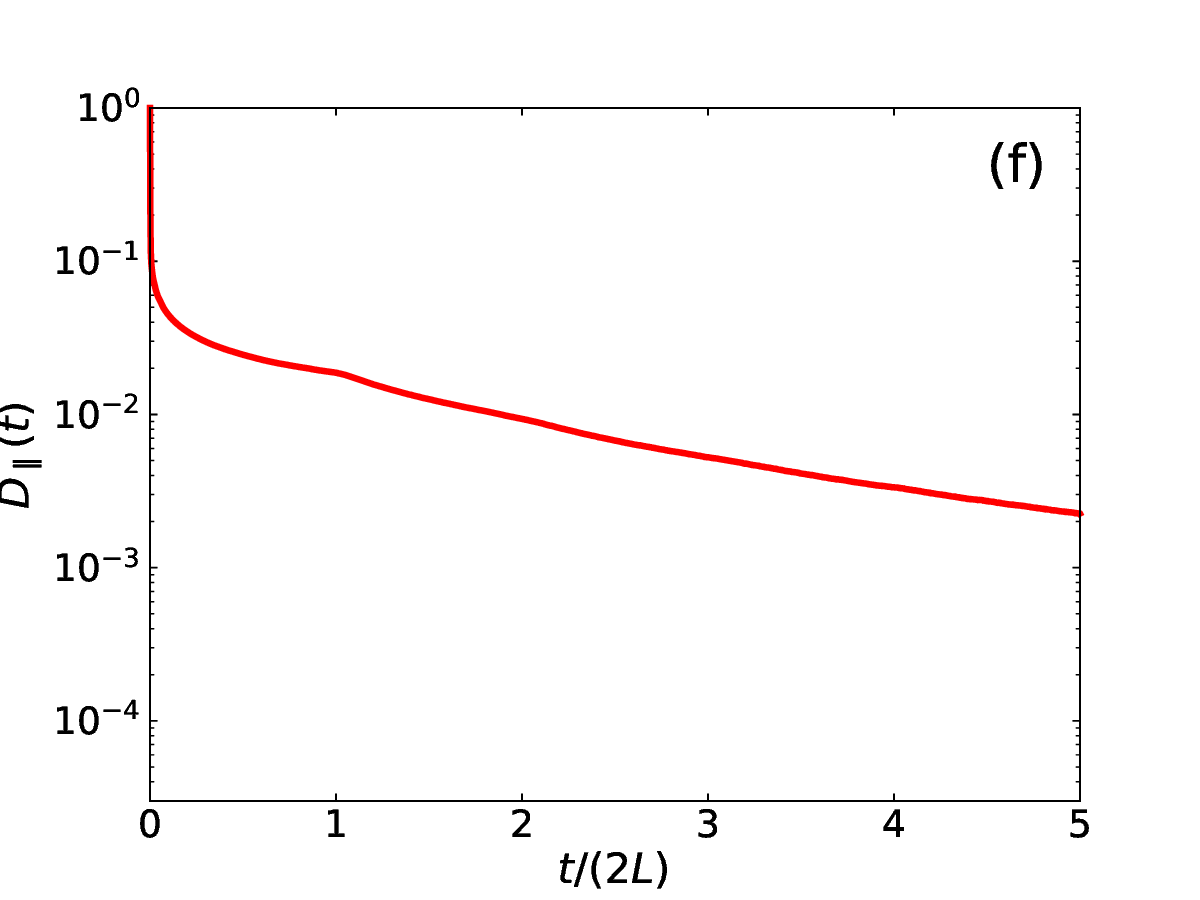} 
\caption{
(Color online)
Simulation results for the distance measure $D_{\parallel}(t)$,
given by Eq. (\ref{eq:D_parallel_n}), as a function of the time $t/(2L)$
(in units of cycles)
on a lattice of size $L=1,024$ (left column) and $L=4,096$ (right column)
and densities (a,d) $p=0.05$; (b,e) $p=0.15$ and (c,f) $p=0.25$.
It is found that starting from a random initial state, $D_{\parallel}(t)$
sharply decays (by more than an order of magnitude)
within a few time steps and then continues to 
decay as time evolves.
It eventually converges toward $D_{\parallel}(t) = 0$.
The results for the two lattice sizes are very similar, 
indicating that the dependence on the system size is weak.
Data points are shown for all the even time steps.
Each data point represents an average over 100 random initial conditions.
The noise in the tail is due to the fact that at these long times the
signal is very small and the fluctuations become significant.
The noise level tends to decrease as the lattice size is increased.
}
\label{fig:2}
\end{figure}

In Fig. \ref{fig:2} we present simulation results  
for the distance measure $D_{\parallel}(t)$,
given by Eq. (\ref{eq:D_parallel_n}), as a function of the time $t/(2L)$ 
(in units of cycles)
for the BML traffic flow model
on lattices of size $L=1,024$ (left column) and $L=4,096$ (right column)
and densities (a,d) $p=0.05$; (b,e) $p=0.15$ and (c,f) $p=0.25$.
It is found that starting from a random initial state, $D_{\parallel}(t)$
sharply decays (by about an order of magnitude)
within a few time steps.
It then continues to 
decay more slowly as time evolves, until it
eventually converges toward $D_{\parallel} = 0$. 
The extremely fast decay of $D_{\parallel}(t)$ during the first few
time steps resembles the behavior of the one-dimensional
system (CA 184), consisting of a single row (with H-cars) or a single 
column (with V-cars), in the low-density limit
\cite{Jha2025}.
During the first few steps it decays like
$D_{\parallel}(t) \propto p^t$ 
or
$D_{\parallel}(t) \propto e^{- t/\tau_{\parallel}}$,
where $\tau_{\parallel} \sim 1 / \ln (1/p) < 1$.
It appears that the initial fast decay of $D_{\parallel}(t)$ 
is a local process that depends on the density $p$ and
does not depend on the lattice size $L$.
At later times, the slower decay of $D_{\parallel}$ is a secondary
effect of the delays due to interactions between cars of different types.
Essentially, an H (V) car that is blocked by a V (H) car may delay another H (V)
car that comes from behind.
Thus, apart from the first few time steps
$D_{\parallel}(t)$ may be considered as a fast variable whose time dependence
is determined by the evolution of $D_{\perp}(t)$, which is a slow variable
\cite{Kampen1985}.

\begin{figure}
\includegraphics[width=7.5cm]{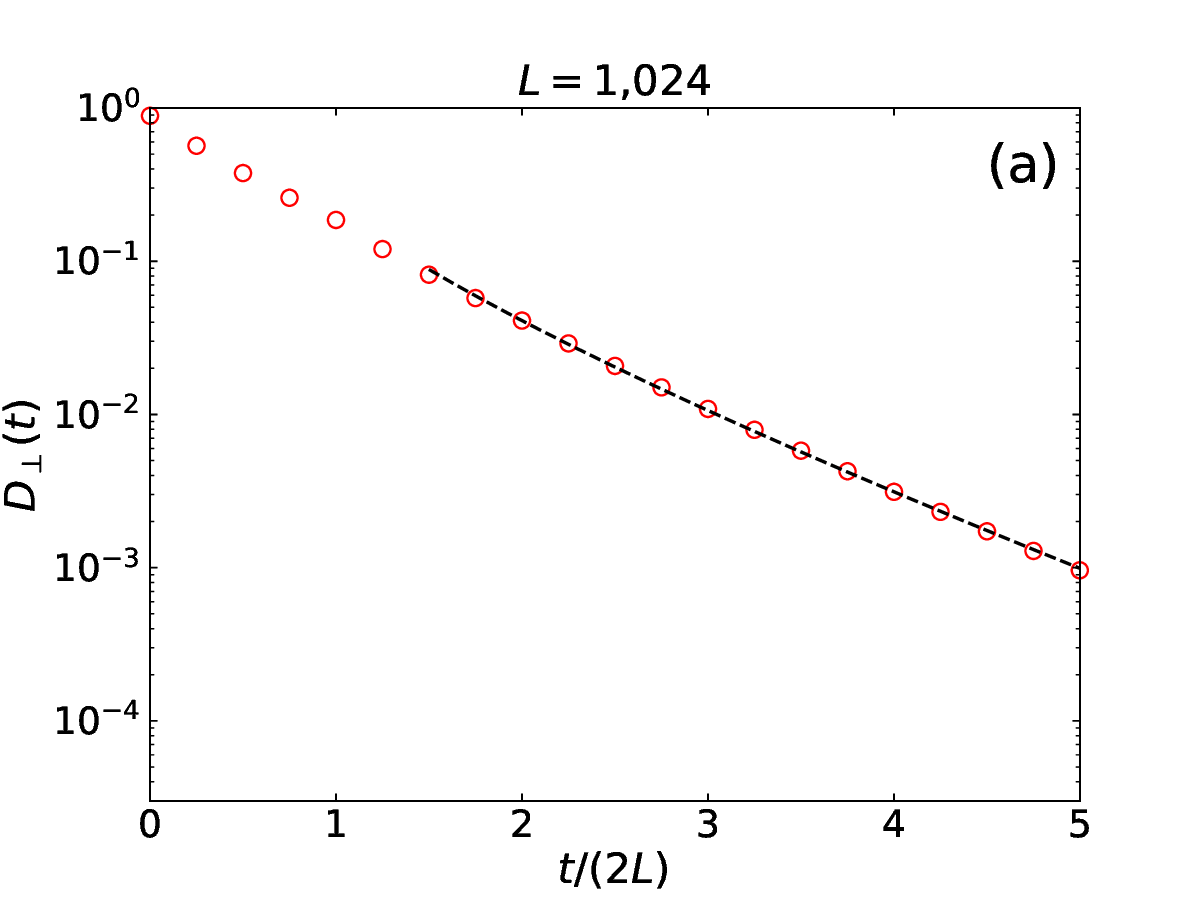} 
\includegraphics[width=7.5cm]{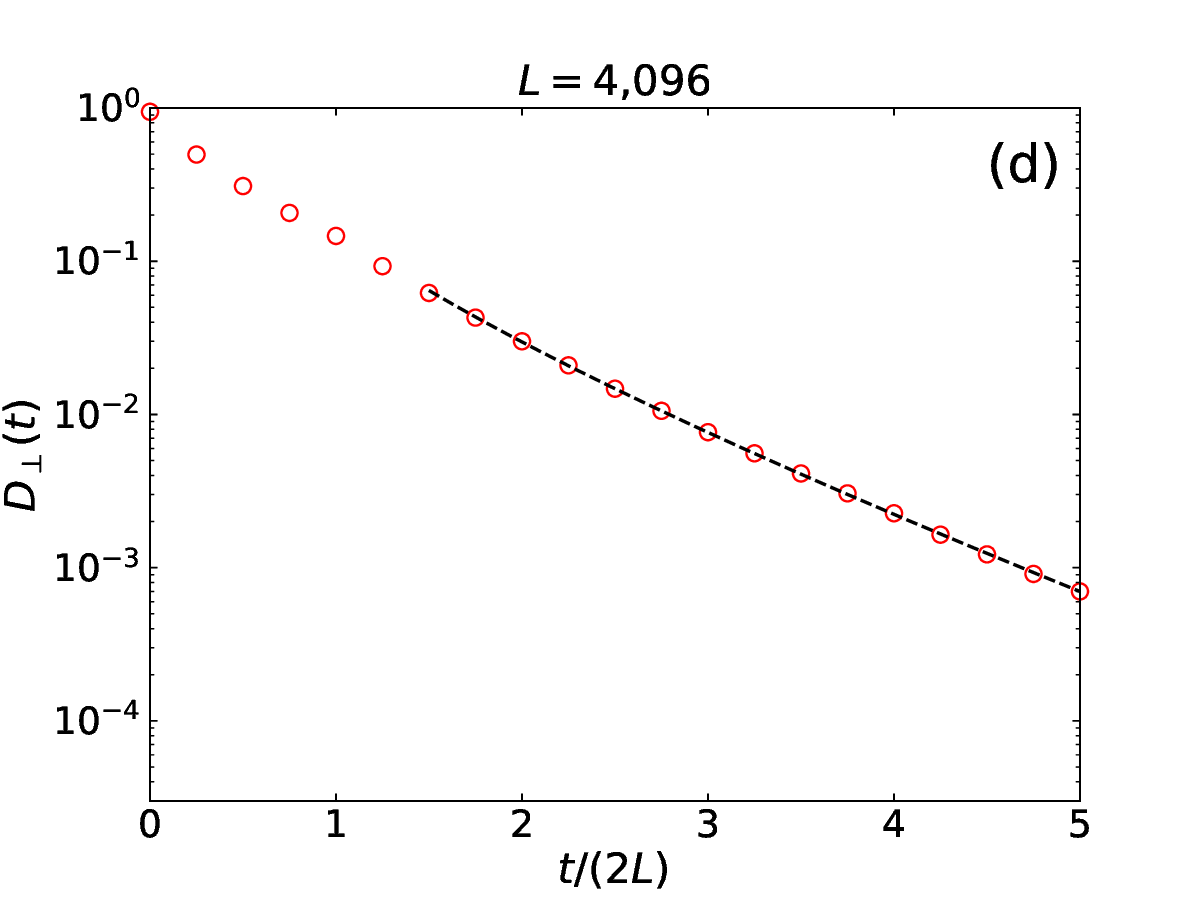} 
\\
\includegraphics[width=7.5cm]{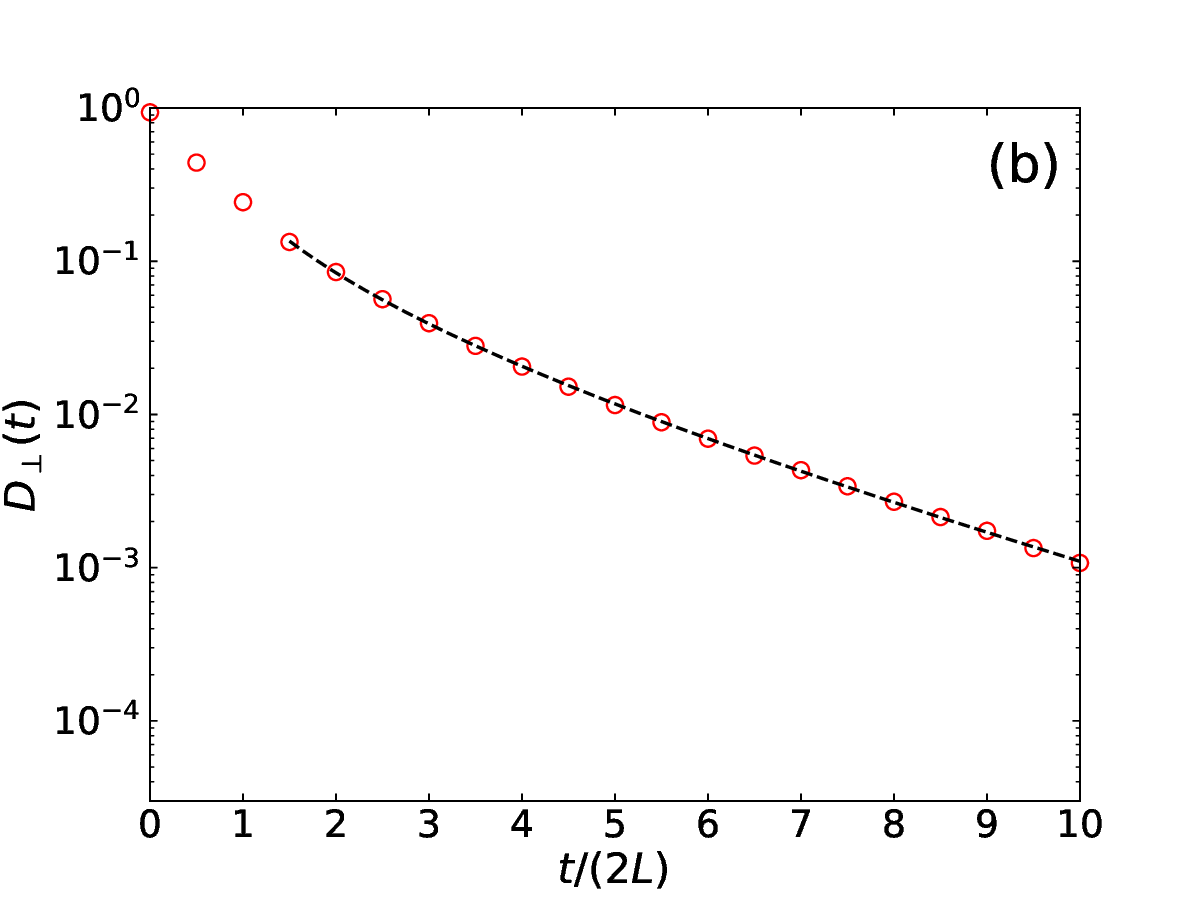} 
\includegraphics[width=7.5cm]{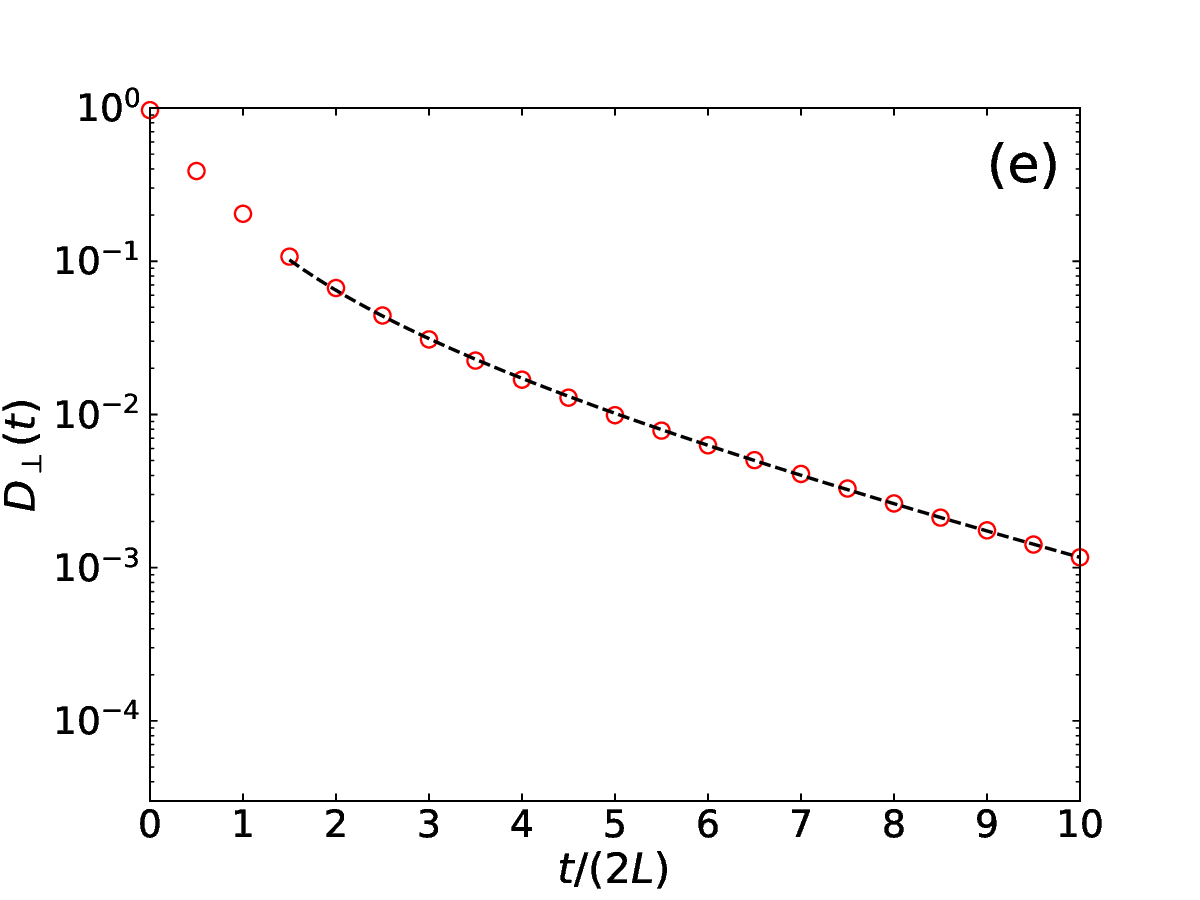} 
\\
\includegraphics[width=7.5cm]{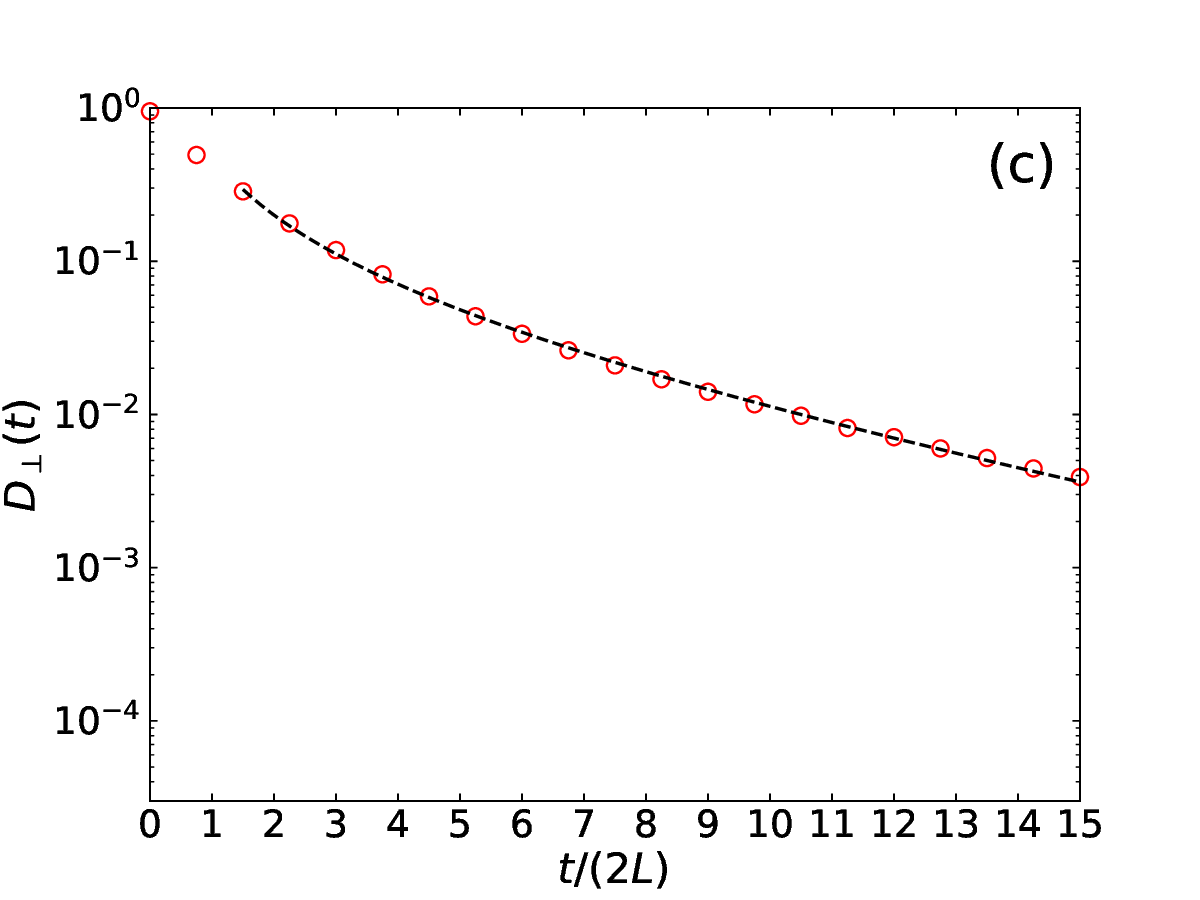} 
\includegraphics[width=7.5cm]{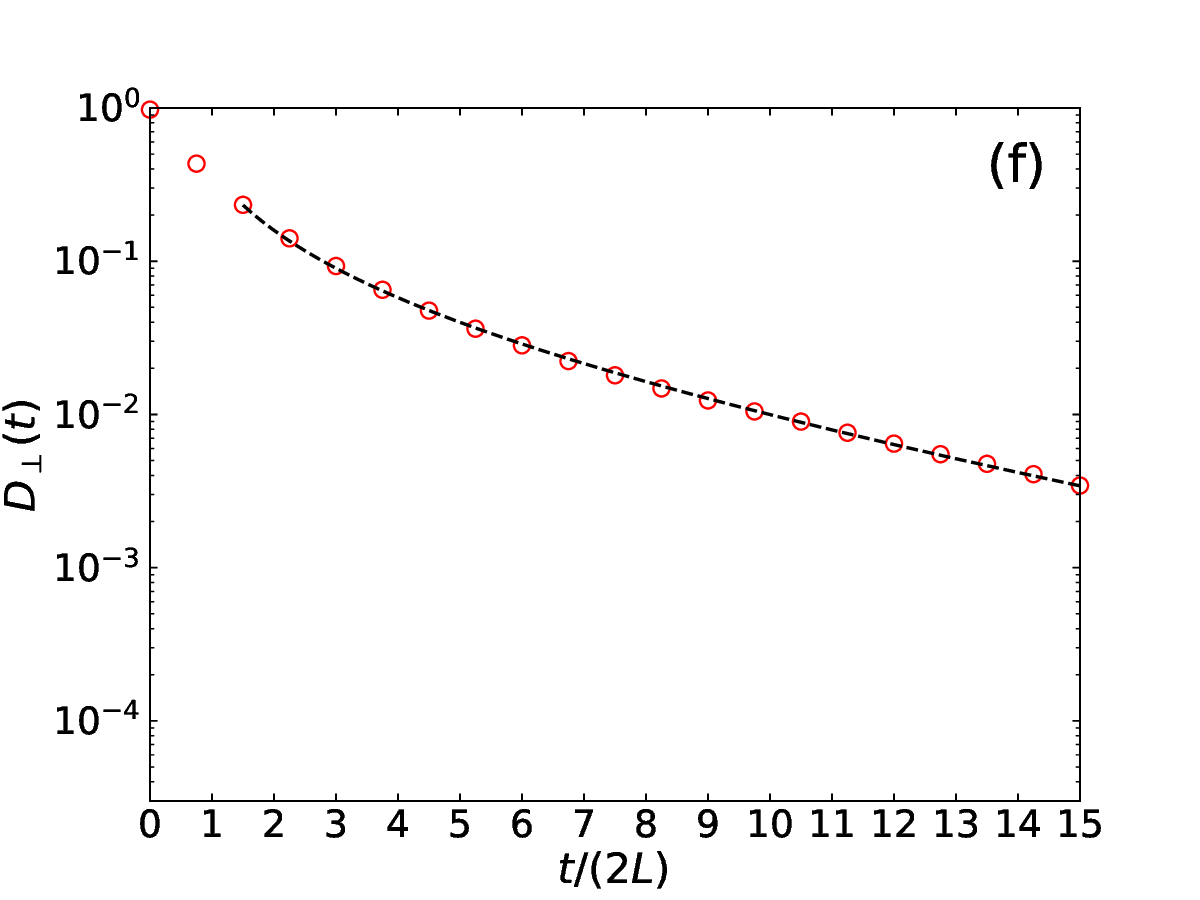} 
\caption{
(Color online)
Simulation results (circles) for the distance measure $D_{\perp}(t)$, 
given by Eq. (\ref{eq:D_perp_n}), as a function of the time $t/(2L)$
(in units of cycles)
for lattices of size $L=1,024$ (left column) and
$L=4,096$ (right column) and densities (a,d) $p=0.05$; (b,e) $p=0.15$ and (c,f) $p=0.25$.
Each data point represents an average over 100 random initial conditions.
The results are well fitted by exponentially truncated power-law decay functions
(dashed lines)
of the form
$D_{\perp}(t) \sim t^{-\gamma} \exp ( - t/\tau_{\perp} )$,
with $\gamma = 1.1$ in all the cases
and the relaxation times (in units of cycles)  
$\tau_{\perp}/(2L)$ take the values
(a) 1.1; (b) 3.1; (c) 7.2; (d) 1.1; (e) 3.6 and (f) 8.0.
}
\label{fig:3}
\end{figure}

In Fig. \ref{fig:3} we present simulation results (circles) 
for the distance measure $D_{\perp}(t)$,
given by Eq. (\ref{eq:D_perp_n}), as a function of the time $t/(2L)$ 
(in units of cycles)
for the BML traffic flow model
on lattices of size $L=1,024$ (left column) and $L=4,096$ (right column)
and densities (a,d) $p=0.05$; (b,e) $p=0.15$ and (c,f) $p=0.25$.
The results are well fitted by an exponentially truncated power-law functions (dashed lines) 
of the form

\begin{equation}
D_{\perp}(t) \sim t^{-\gamma} \exp ( - t/\tau_{\perp} ),  
\end{equation}

\noindent
where the exponent takes the value $\gamma = 1.1$ in all the cases
and the relaxation times  
$\tau_{\perp}/(2L)$ 
(in units of cycles)
take the values
(a) 1.1; (b) 3.1; (c) 7.2; (d) 1.1; (e) 3.6 and (f) 8.0.
Such power-law decay is common in coarsening processes that
are driven by the interaction between pairs of particles
\cite{Toussaint1983,Kang1984,Bramson1988}.
In unary systems that consist of a single particle species 
the power-law decay may last for many orders of magnitude
\cite{Toussaint1983}.
In contrast, in binary systems that consist of two particle species
that interact with each other, the power-law decay is often truncated
due to emergence of segregation between the two species
\cite{Kang1984,Bramson1988}.

\begin{figure}
\includegraphics[width=9.0cm]{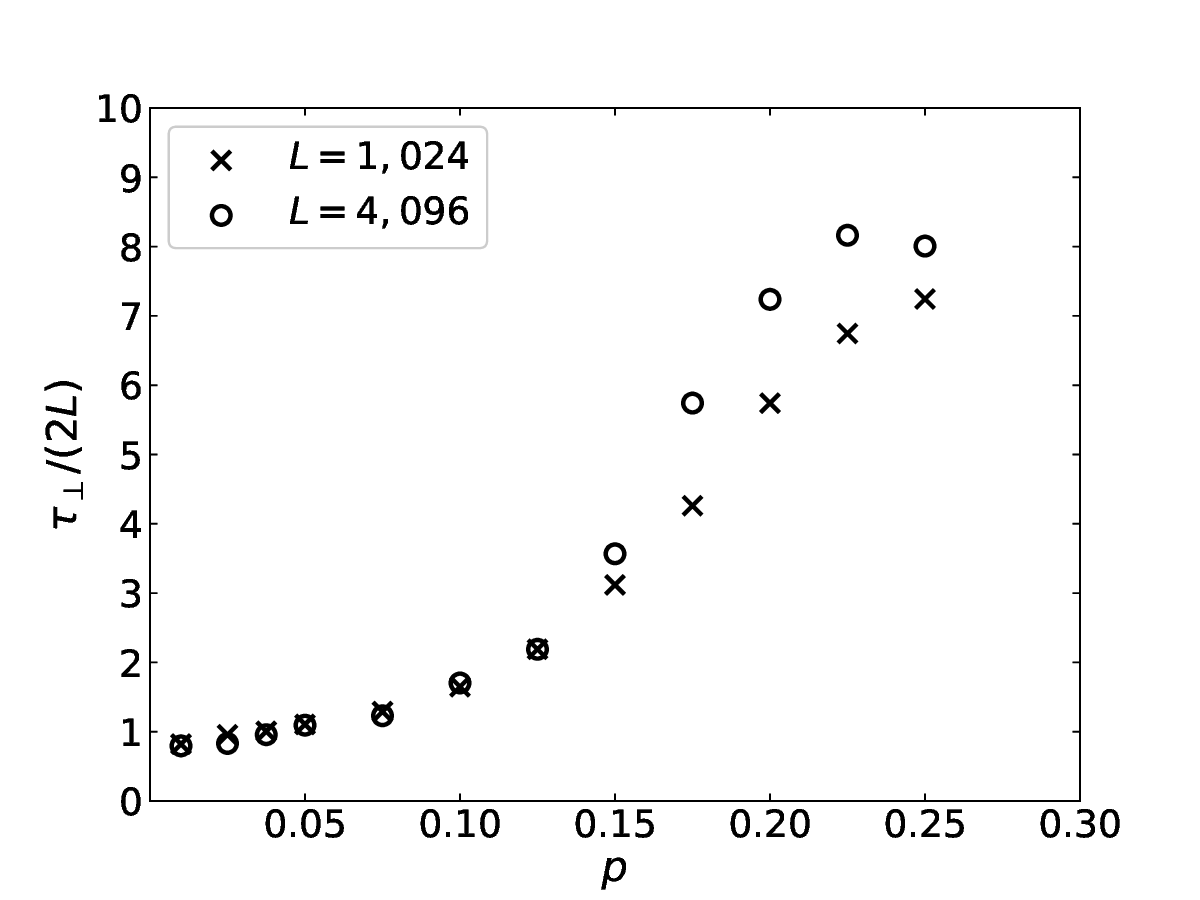} 
\caption{
The relaxation time $\tau_{\perp}/(2L)$ 
(in units of cycles)
of the distance measure $D_{\perp}(t)$ 
as a function of the density $p$, for lattices of size
$L=1,024$ ($\times$) and $L=4,096$ ($\circ$).
In both cases $\tau_{\perp}$ increases as $p$ is increased and then starts to saturate at $p=0.25$
(and even slightly decreases for $L=4,096$). 
The results indicate that for $p > 0.125$ the relaxation time tends to increase as
the system size is increased.
For $L=1,024$ each data point represent an average over $1,000$ random initial conditions,
while for $L=4,096$ each data point represents an average over $100$ random instances.
}
\label{fig:4}
\end{figure}

In Fig. \ref{fig:4} we present 
the relaxation time $\tau_{\perp}$ of the distance measure $D_{\perp}(t)$ as a function of the
density $p$ for the BML traffic model on lattices of size 
$L=1,024$ ($\times$)
and
$L=4,096$ ($\circ$).
For low densities in the range of $p \le 0.125$
the results obtained for these two system sizes essentially overlap.
This indicates that for $p \le 0.125$  
the relaxation time $\tau_{\perp}(L)$
scales linearly with the lattice size,
namely
$\tau_{\perp}(L) \propto L$.
In contrast, for
$p > 0.125$ 
the two curves split indicating that in this regime the relaxation time grows 
faster than linearly with the lattice size.

Another feature observed in Fig. \ref{fig:4} is that for both values of $L$
the relaxation time $\tau_{\perp}$ increases as the density 
$p$ is increased up to the data point at
$p = 0.225$.
In the next data point, at $p=0.25$, for the larger lattice of $L=4,096$,
there is a saturation and even a slight decrease of $\tau_{\perp}(L)$.
Additional data points obtained for $p=0.275$ indicate that the saturation
persists above $p=0.25$. However, the quality of the fit deteriorates at
$p=0.275$ and therefore these data points are not shown.

Saturation behavior is common in the vicinity of weakly first-order transitions.
In these systems, the correlation length and the correlation time 
grow as the transition is approached
(as in second-order transitions, in which they diverge),
but then they saturate very close to the transition point
\cite{Gavai1989,Buffenoir1993}.

Careful inspection of simulation movies for system sizes
of $L=512$ and larger reveal that there is an apparent
qualitative change in the dynamics around $p=0.25$.
For $0 < p < 0.25$ the segregation process is slow and is driven by local
rearrangements involving a small number of cars at a time.
For $0.25 \le p < 0.3$ large scale avalanches emerge that lead to fast
coordinated rearrangements of cars across multiple diagonal bands.
These avalanches apparently expedite the convergence to FFP states
and shorten the relaxation time.
At even higher densities, above $p_c \simeq 0.3$, these avalanches  
eventually destabilize the free-flowing states and drive the system
toward fully-jammed states or intermediate states of congested traffic.

The simulations presented in this paper were carried out 
on an NVIDIA RTX 2000 Ada GPU with 16GB of memory.
The computer codes were written in Python using the Numba library.

\section{Discussion}

The BML model exhibits a first order dynamical phase transition between the free-flowing phase 
and the fully-jammed phase
\cite{Biham1992}. 
This transition occurs at a critical density $p_c(L)$ that depends on the system size. 
For sufficiently large systems (roughly $L \ge 512$) the transition appears to be first-order in nature, 
marked by a discontinuous jump in the average velocity
\cite{Biham1992}. 
For a system size of $L=512$ the transition density is approximately
$p_c(L) \simeq 0.31$
\cite{Biham1992}. 
As the system size is increased it tends to decrease in a very slow fashion.
In a narrow region just above the critical density, the system may evolve into long-lived 
congested states in which domains of jammed traffic and free-flowing traffic coexist
\cite{Dsouza2005,Olmos2015}.

Deterministic models such as the BML model provide a baseline for a range
of stochastic models of traffic flow.
Stochastic effects are intrinsic to traffic modeling due to the unperdictability
of human behavior as well as technical and environmental variabilities.
There are several stochastic variants
of the BML model
\cite{Cuesta1993,Molera1995,Ding2011}.
In the stochastic model of Refs.
\cite{Cuesta1993,Molera1995},
at each time step an H-car may move like a V-car
with probability $r < 1/2$, while a V-car 
can move like an H-car with the same probability.
In the low-density phase, the mean velocity gradually decreases
as $p$ is increased, and thus the model does not converge
to an FFP state. At the threshold density $p_c(r)$
the system exhibits a sharp transition
from the predominantly flowing 
phase to a predominantly jammed phase.

In the stochastic variant studied in Refs.
\cite{Ding2011},
each car may skip a move with probability $r$ even in
case that the cell in front of it is empty.
This model still exhibits a sharp dynamical transition
between a flowing phase and a fully jammed phase.
However, the flowing phase is repeatedly randomized
and does not converge toward an FFP state.
The mean velocity in this phase decreases as
$p$ is increased in a way that can be qualitatively
accounted for by a mean-field analysis
\cite{Ding2011}.
In this variant the intermediate phase is washed out.

Two-dimensional traffic flow models with open boundaries were also studied
\cite{Tadaki1996,Cividini2013}.
In these models cars are injected into the system from the top and left
boundaries and leave the system from the bottom and right boundaries.
It was found that even in the limit of low feeding rate the system does not
converge into a free-flowing state 
\cite{Tadaki1996,Cividini2013}. 
This is due to the fluctuations induced 
by the random feeding process.

Apart from computer simulations, the phase transition in the BML model was
also studied using a mean-field approach
\cite{Molera1995,Wang1996,Ding2013}.
Using self-consistent equations, Wang et al. obtained an approximate result for
the critical density, at $p_c \simeq 0.34$, which is slightly above the value
obtained in computer simulations
\cite{Wang1996}.
The mean-field approach is suitable for the stochastic variants of 
BML model and for variants which employ open boundaries.
It is not expected to be valid in the case of periodic boundary conditions
in which strong correlations build up.

\section{Summary}

We analyzed the structure and dynamics of 
the states of free-flowing traffic that emerge 
in the low-density phase
of the BML traffic flow model.
To this end
we introduced the distance measure 
$D(t) = D_{\parallel}(t) + D_{\perp}(t)$
in the configuration space
between the state of the system at time $t$
and the set of free-flowing states.
The term $D_{\parallel}(t)$ accounts for the interactions between cars
of the same type, while $D_{\perp}(t)$ accounts for the interactions 
between cars of different types.
We showed that for the free-flowing periodic states $D(t)=0$, while for all the other states $D(t) > 0$.
We analyzed the time dependence of $D_{\parallel}(t)$ and $D_{\perp}(t)$ 
during the convergence toward the FFP states and 
found that there is a separation of time scales, where $D_{\parallel}(t)$ decays
within a few time steps while $D_{\perp}(t)$ decays much more slowly.
It is also shown that the time dependence of $D_{\perp}(t)$ is well fitted by
an exponentially truncated power-law decay of the form
$D_{\perp}(t) \sim t^{-\gamma} \exp ( - t/\tau_{\perp} )$,
in which the relaxation time $\tau_{\perp}$ depends on $L$ and $p$ 
while the power-law exponent   
$\gamma \simeq 1.1$ is fixed.
The power-law decay suggests avalanche-like dynamics with no characteristic scale,
while the exponential cutoff is imposed by the finite lattice size.

The distance measures $D_{\parallel}(t)$ and $D_{\perp}(t)$ and
the separation of time scales between them provide useful insight for the development
of a theory for the dynamics of the BML traffic flow model in the low-density phase. 
These results indicate that such a theory should focus on the interactions between pairs of cars of 
different types, which are the slow variables in the dynamics. The interactions between pairs of
cars of the same type are fast variables 
\cite{Kampen1985},
which relax almost instantaneously.
More specifically, these distance measures may be helpful in the formulation of coarse-grained descriptions
and mean-field equations for the dynamics of the model.
In such mean-field equations, the distance measure $D_{\perp}(t)$ may be cast in the form of a
Lyapunov function
\cite{Giesl2015},
which decreases monotonically during the relaxation process.
It will also be interesting to extend the analysis to stochastic variants of the BML model,
where the distance measures may be used to quantify the stability and robustness of
the free-flowing states.

\appendix
\section{The consistency of the evolution equations}

Below we show some important properties of 
the evolution equations (\ref{eq:Hupdate}) and (\ref{eq:Vupdate}).
For concreteness we take the case in which
$t=2s$ such that in the next time step it is the
turn of the H-cars to move.
Below we show that if $H_{2s}(i,j)=0$ or $1$
and $V_{2s}(i,j)=0$ or $1$
and $H_{2s}(i,j) V_{2s}(i,j) = 0$
for all values of $i,j=0,1,\dots,L-1$,
then
$H_{2s+1}(i,j) = 0$ or $1$
for all values of $i$ and $j$.

The condition
$H_{2s}(i,j) V_{2s}(i,j) = 0$
implies that
$H_{2s}(i,j) + V_{2s}(i,j) = 0$ or $1$
for all values of $i,j=0,1,\dots,L-1$.
From Eq. (\ref{eq:Hupdate}) it is found that if
$H_{2s}(i,j) + V_{2s}(i,j) = 0$, then
$H_{2s+1}(i,j) = H_{2s}(i,j-1)$.
Otherwise, namely in case that
$H_{2s}(i,j) + V_{2s}(i,j) = 1$,
there are two possibilities:
if $H_{2s}(i,j+1) + V_{2s}(i,j+1) = 1$
then $H_{2s+1}(i,j) = H_{2s}(i,j)$,
otherwise $H_{2s+1}(i,j) = 0$.
In all these possibilities we find that $H_{2s+1}(i,j)$ is
either $0$ or $1$.
This confirms the closure of 
the evolution equation (\ref{eq:Hupdate}) over the field of
binary numbers.
A similar argument applies also to Eq. (\ref{eq:Vupdate}).

It is also important to show that if the condition of Eq. (\ref{eq:no_overlap}) is
satisfied at time $t=2s$, namely if
$H_{2s}(i,j) V_{2s}(i,j) = 0$,
then it is satisfied at time $t=2s+1$ as well, namely

\begin{equation}
H_{2s+1}(i,j) V_{2s+1}(i,j) = 0.
\label{eq:HV_2sp1}
\end{equation}

\noindent
Inserting  $H_{2s+1}(i,j)$ from Eq. (\ref{eq:Hupdate}) into Eq. (\ref{eq:HV_2sp1}),
it is found that all the terms include products that vanish,
thus confirming the validity of Eq. (\ref{eq:HV_2sp1}).

\section{Dynamical stability of the FFP state}

Below we show that if
$d_{\parallel}(t)=d_{\perp}(t)=0$ 
then
$d_{\parallel}(t+1)=d_{\perp}(t+1)=0$.
For concreteness we take the case in which
$t=2s$ such that in the next time step it is the
turn of the H-cars to move.
The condition that $d_{\parallel}(2s)=0$
implies that

\begin{equation}
H_{2s}(i,j) H_{2s}(i,j+1) = 0
\label{eq:HH0}
\end{equation}

\noindent
and

\begin{equation}
V_{2s}(i,j) V_{2s}(i+1,j) = 0
\label{eq:VV0}
\end{equation}

\noindent
for all values of $i,j = 0,1,\dots,L-1$.
Similarly, one of the implications of the condition that $d_{\perp}(2s) = 0$ 
is that

\begin{equation}
H_{2s}(i,j) V_{2s}(i,j+1) = 0
\label{eq:HV0}
\end{equation}
 
\noindent
for all values of $i,j = 0,1,\dots,L-1$.
 
Assuming that
$d_{\parallel}(2s) = d_{\perp}(2s) = 0$
we now show that
$d_{\parallel}(2s+1) = 0$.
Clearly, the second term on the right hand side of Eq. (\ref{eq:D_par})
remains unchanged between $t=2s$ and $t=2s+1$, and we thus need to 
focus on the first term.
To evaluate the sum

\begin{equation}
\sum_{i,j=0}^{L-1} H_{2s+1}(i,j) H_{2s+1}(i,j+1)
\end{equation}
 
\noindent
we express both factors in the product on the right hand side
using the evolution equation (\ref{eq:Hupdate}).
Carrying out the multiplication, one obtains a sum of terms such 
that each term includes a product of the form given in Eqs.
(\ref{eq:HH0}) or (\ref{eq:HV0}).
This implies that all these terms vanish, proving that
$d_{\parallel}(2s+1) = 0$.

Assuming that
$d_{\parallel}(2s) = d_{\perp}(2s) = 0$
we now show that
$d_{\perp}(2s+1) = 0$.
To this end, we write $d_{\perp}(2s+1)$ explicitly
in the form

\begin{equation}
d_{\perp}(2s+1) = 
\sum_{n=0}^{L-1} \min \{ h_{2s+1}(n),v_{2s+1}(n) \}
+
\sum_{n=0}^{L-1} \min \{ v_{2s+1}(n),h_{2s+1}(n+1) \}.
\label{eq:D_perpApp}
\end{equation}
 
\noindent
Using the definition of $v_t(n)$, given by Eq. (\ref{eq:v_t_n}) and the fact that
$V_t(i,j)$ does not change in the odd time steps 
[Eq. (\ref{eq:Hupdate})],
we conclude that

\begin{equation}
v_{2s+1}(n) = v_{2s}(n).
\label{eq:v_2sp1}
\end{equation}

\noindent
Under the conditions specified above, 
at $t=2s+1$ all the H-cars move without 
obstruction, which implies that

\begin{equation}
H_{2s+1}(i,j) = H_{2s}(i,j-1).
\label{eq:H2sp1}
\end{equation}

\noindent
Combining this result with the definition of $h_t(n)$,
given by Eq. (\ref{eq:h_t_n}), we obtain

\begin{equation}
h_{2s+1}(n) = h_{2s}(n-1).
\label{eq:h_2sp1}
\end{equation}

\noindent
Inserting $v_{2s+1}(n)$ from Eq. (\ref{eq:v_2sp1})
and $h_{2s+1}(n)$ from Eq. (\ref{eq:h_2sp1})
into Eq. (\ref{eq:D_perpApp}), we obtain

\begin{equation}
d_{\perp}(2s+1) 
%= d_{\perp}(2s) 
= 0.
\end{equation}


\begin{thebibliography}{10}
\bibitem{Hamming}
The Hamming distance in a discrete $L$-dimensional vector space
between two vectors 
$\vec x = (x_1,x_2,\dots,x_L)$ 
and 
$\vec y = (y_1,y_2,\dots,y_L)$
is given by
$d(x,y) = \sum_{i=1}^{L} | y_i - x_i |$,
where $|x|$ is the absolute value of $x$.
\end{thebibliography}
\end{document}